\documentclass{article}
\usepackage{spconf,amsmath,graphicx, amsfonts}
\usepackage{multicol,multirow}
\usepackage{url,cite}
\usepackage{hyperref}
\usepackage[table]{xcolor}
\usepackage{xcolor}
\usepackage{booktabs, makecell}
\usepackage{enumitem}
\setlist[itemize]{align=parleft,left=0pt..1.3em}
\usepackage{colortbl}
\usepackage{tabularx,lipsum}
\hypersetup{
    colorlinks=true,
    linkcolor=purple,
    filecolor=magenta,      
    urlcolor=purple,
}
\newcommand{\newpara}[1]{\vspace{8pt}\noindent\textbf{#1}}
\newcolumntype{Y}{>{\centering\arraybackslash}X}
\usepackage{caption}
\usepackage{subcaption}

\makeatletter
\def\ps@IEEEtitlepagestyle{%
\def\@oddfoot{\mycopyrightnotice}%
\def\@evenfoot{}%
}
\def\mycopyrightnotice{%
{\footnotesize 978-1-6654-7189-3/22\$31.00~\copyright~2023 IEEE\hfill} % MODIFY this line according to the code! 
\gdef\mycopyrightnotice{}
}
\title{Frequency and Multi-Scale Selective Kernel Attention\\for Speaker Verification}

\name{Sung Hwan Mun$^*$$^1$ \,\, Jee-weon Jung$^*$$^2$ \,\, Min Hyun Han$^1$ \,\, Nam Soo Kim$^1$
  \thanks{* Equal contribution.}
  \thanks{This work was supported by Institute of Information \& communications Technology Planning \& Evaluation (IITP) grant funded by the Korea government (MSIT) (No.2021-0-00456, Development of Ultra-high Speech Quality  Technology for Remote Multi-speaker Conference System).}
}
\address{
  $^1$Department of ECE and INMC, Seoul National University, Seoul, South Korea \\
  $^2$Naver Corporation, South Korea \\
}
\begin{document}
\ps@IEEEtitlepagestyle

\maketitle

\begin{abstract}
The majority of recent state-of-the-art speaker verification architectures adopt multi-scale processing and frequency-channel attention mechanisms.
Convolutional layers of these models typically have a fixed kernel size, e.g., 3 or 5.
In this study, we further contribute to this line of research utilising a selective kernel attention (SKA) mechanism. 
The SKA mechanism allows each convolutional layer to adaptively select the kernel size in a data-driven fashion. 
It is based on an attention mechanism which exploits both frequency and channel domain.
We first apply existing SKA module to our baseline.
Then we propose two SKA variants where the first variant is applied in front of the ECAPA-TDNN model and the other is combined with the Res2net backbone block.
Through extensive experiments, we demonstrate that our two proposed SKA variants  consistently improves the performance and are complementary when tested on three different evaluation protocols.
%Implementation is available at \small{\url{https://github.com/msh9184/ska-tdnn.git}}.
\end{abstract}
\begin{keywords}
speaker verification, selective kernel attention, multi-scale module
\end{keywords}

\section{Introduction}
In recent years, various deep neural network (DNN) architectures for speaker verification (SV) systems have been proposed~\cite{wan2018generalized, jung2019RawNet, vox1, vox2, chung2020in, heo2020clova}.
Current state-of-the-art architectures typically utilise 1-dimensional convolutional neural networks (1D-CNNs) such as x-vector, RawNet3, or ECAPA-TDNN~\cite{snyder2018x,jung2022pushing,desplanques2020ecapa}.
Among these, ECAPA-TDNN~\cite{desplanques2020ecapa} is widely adopted, demonstrating stable yet competitive performance across a wide range of studies.
It involves Res2net backbone blocks with a squeeze-excitation (SE) layer at the end of each block, where the Res2net incorporates multi-scale modelling and the SE efficiently recalibrates the channel (filter) axis of a CNN feature map~\cite{hu2018squeeze, gao2019res2net}. 

Several architectures that extend ECAPA-TDNN have also been proposed~\cite{thienpondt2021ecapacnn, liu2022mfa}.
Authors of \cite{thienpondt2021ecapacnn} extended ECAPA-TDNN and proposed ECAPA-CNN-TDNN, by adding a 2D-CNN-based front-end with frequency-wise SE layers to incorporate frequency translational invariance.
Similarly, MFA-TDNN~\cite{liu2022mfa} applied a 2D-CNN-based module in front of the original ECAPA-TDNN identical to the ECAPA-CNN-TDNN; however, it proposed to replace the 2D-CNN-based module with a multi-scale frequency-channel attention module.
Leveraging the multi-scale processing capability and the attention module, which resembles the SE layer, MFA-TDNN demonstrates competitive performance across test scenarios involving diverse duration.

To this end, we explore to further push this line of research. % 수정 2022.07.19.
We adapt the selective kernel attention (SKA) mechanism more effectively to speaker verification, inspired by \cite{li2019selective, zhang2020resnest, chen2020dynamic, wu2021rsknet, kim2022decomposed}.
Speech signals have multi-scale and hierarchical linguistic structures (e.g., phoneme, syllable, and word) and different time-frequency responses~\cite{zhang2022tms}.
The SKA module is expected to adaptively emphasise the local and global information required for extracting robust speaker-discriminative representations.
Hence, our model architecture, which involves several different-sized kernels can choose which kernel to concentrate on in a \emph{data-driven} fashion.

We further propose two modules, which are variants of the SKA.
First, we propose multi-scale SKA (msSKA) which incorporates the SKA approach with the Res2net-based backbone modules.
The objective is to develop a backbone module which can better model utterances with diverse durations.
Second, we propose frequency-wise SKA (fwSKA) which adapts the SKA module to operate upon the frequency axis of a feature map.
It is designed to inject global frequency information across the intermediate feature representations, similarly to~\cite{thienpondt2021ecapacnn}. 

Experiments conducted with three different evaluation protocols consistently demonstrate the effectiveness of our proposed approaches over the baseline systems. 
We also observe the identical tendency across three different durations.

The rest of this paper is organised as follows: Section~\ref{ssec:cwSKA} describes the selective kernel attention module, Section~\ref{ssec:ska-variants} introduces the proposed SKA-variants, and Sections~\ref{ssec:model} presents the proposed architectures. Then the experimental settings and results are addressed in Sections~\ref{ssec:exp} and ~\ref{ssec:result}, respectively. Finally, we conclude in Sections~\ref{ssec:conclude}.
%The rest of this paper is organised as follows: Section 2 describes the selective kernel attention module, and Section 3 presents the proposed architectures. Then, the experiments and results are addressed in Sections 4 and 5, respectively. Finally, we conclude in Section 6.

\begin{figure}[t!]
\begin{minipage}[b]{1.0\linewidth}
  \centering
  \centerline{\includegraphics[width=8.3cm]{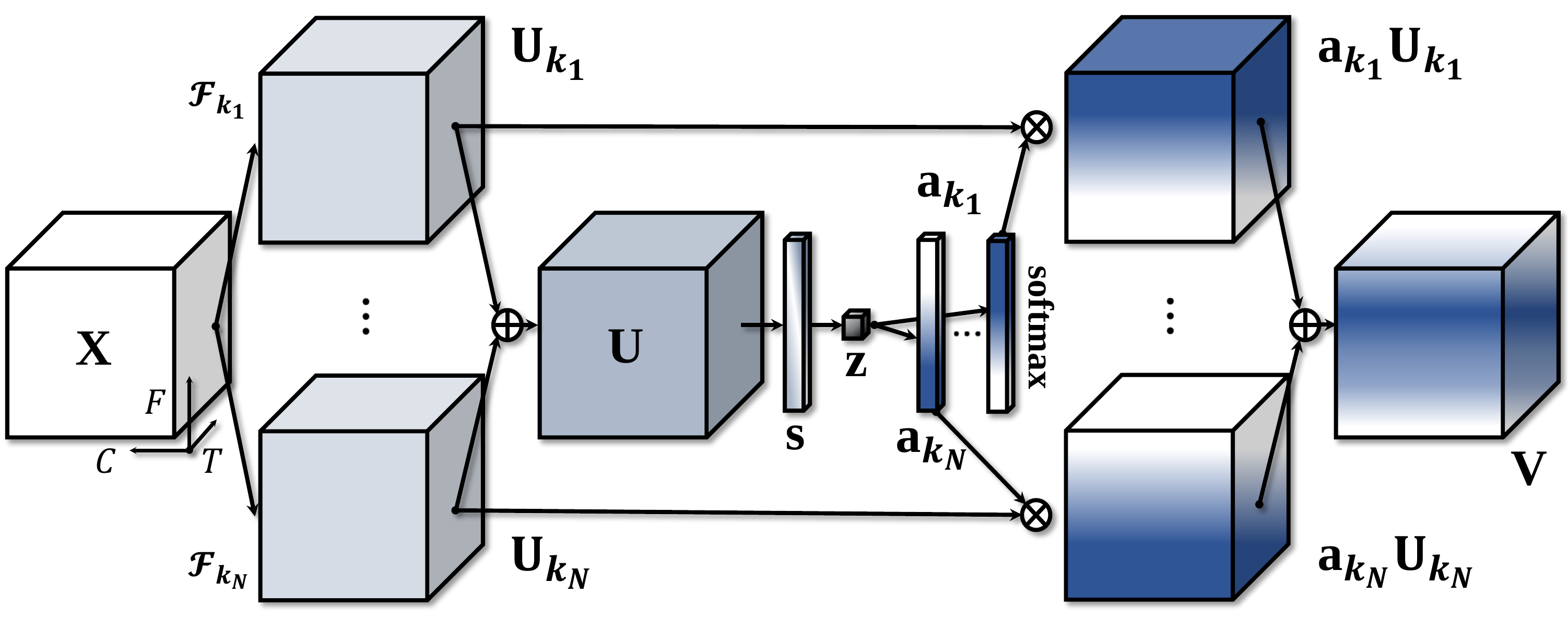}}
\vspace{0cm}
\end{minipage}
\centering
\caption{Frequency-wise selective kernel attention (fwSKA)}
\label{fig:fwSKA}
\end{figure}

\section{Selective kernel attention module}
\label{ssec:cwSKA}
This Section describes the selective kernel attention (SKA) mechanism \cite{li2019selective} which can select the kernel size adaptive in a data-driven fashion.

For a given $\textbf{X}\in\mathbb{R}^{C'\times F'\times T'}$, let $\mathcal{F}_{k_{i}}:\textbf{X} \rightarrow \textbf{U}_{k_{i}}\in\mathbb{R}^{C \times F \times T}$ be a convolution operator with kernel size $k_{i}$.
First, the input feature map $\textbf{X}$ is split into $N$ branches.
Each convolution layer, $\mathcal{F}_{k_{i}}$, generates $\textbf{U}_{k_{i}}$, where each of the $\{k_{i}\}^{N}_{i=1}$ has a pre-defined different kernel size.  
To integrate different scales of information into the next layer, $N$ branches are fused by an element-wise summation, i.e., $\textbf{U} = \sum_{i=1}^{N} \textbf{U}_{k_{i}}$.
Then 2D global average pooling (GAP) embeds the global information into the channel-wise feature vector $\textbf{s} \in \mathbb{R}^{C}$ as follow:
\begin{equation}
  \textbf{s} = {1 \over {F \times T}}\sum_{f=1}^{F}\sum_{t=1}^{T}\textbf{U}(f,t).
\end{equation}
The fully-connected (FC), batch normalisation (BN) and ReLU layers are sequentially passed to squeeze the channel-wise compact feature $\textbf{z} \in \mathbb{R}^{d}$:
\begin{equation}
  \textbf{z} = \text{ReLU}(\text{BN}(\textbf{W} \textbf{s})),
\end{equation}
where $\textbf{W} \in \mathbb{R}^{d \times C}$ denotes the weight matrix of a FC layer and $d$ is the dimensionality of $\textbf{z}$.
%Next, soft attention weights across channels are calculated via a softmax function as follows:
Next, soft attention weights across channels $\textbf{a}_{k_i} = [a_{k_i;1},...,a_{k_i;C}]^T \in \mathbb{R}^{C}$ are calculated via a softmax function as follows:
% vector간 divide 연산?
\begin{equation}
  a_{k_i;j} = {\exp(A_{k_i;j}\textbf{z}) \over {\sum^{N}_{l=1} \exp(A_{k_{l};j}\textbf{z})}},
\end{equation}
%where $\textbf{A}_{k_i} \in \mathbb{R}^{C\times d}$ is a FC weight matrix and $\textbf{a}_{k_i} \in \mathbb{R}^{C}$ denotes the channel-wise soft attention vector for $\textbf{U}_{k_i}$.
where $A_{k_i;j}\in \mathbb{R}^{d}$ is the $j$-th FC weight row vector of $\textbf{A}_{k_i}=[A_{k_i;1}^T,...,A_{k_i;C}^T]^T \in \mathbb{R}^{C\times d}$.

Finally, the output feature map $\textbf{V}\in \mathbb{R}^{C \times F \times T}$ is computed as the weighed summation over the different branches:
\begin{equation}
  V_{j} = \sum_{i=1}^{N} a_{k_i;j} U_{k_i;j}, \>\>\>
  \sum_{i=1}^{N} a_{k_i;j} = 1.
\end{equation}
where $V_{j}$ and $U_{k_i;j} \in \mathbb{R}^{F\times T}$ are the $j$-th components of $\textbf{V}$ and $\textbf{U}_{k_i}$, respectively. We note the conventional SKA as channel-wise SKA (cwSKA) throughout this paper.

\section{Proposed SKA-variants}
\label{ssec:ska-variants}
\subsection{Frequency-wise SKA (fwSKA)}
\label{ssec:fwSKA}
The conventional SKA method, cwSKA, extracts the global information regarding the channel importance by using 2D GAP on the $F \times T$ dimension. 
However, speaker-discriminative information may also exist in the frequency or temporal domain, which the channel-wise recalibration can not effectively capture. 
Thus, we propose frequency-wise SKA (fwSKA) to aggregate global frequency information to the attention weights using the SKA framework.
It adopts the same SKA technique, however, $\textbf{s}$ is a frequency-wise feature vector rather than a channel-wise feature vector:
\begin{equation}
  \textbf{s} = {1 \over {C \times T}}\sum_{c=1}^{C}\sum_{t=1}^{T}\textbf{U}(c,t).
\end{equation}
% where $\textbf{s}_f \in \mathbb{R}^{F \times 1}$ denotes the frequency-wise feature vector.
% where $\textbf{s} \in \mathbb{R}^{F}$ denotes the frequency-wise feature vector.

\begin{figure}[t!]
\begin{minipage}[b]{1.0\linewidth}
  \centering
  \centerline{\includegraphics[width=8.5cm]{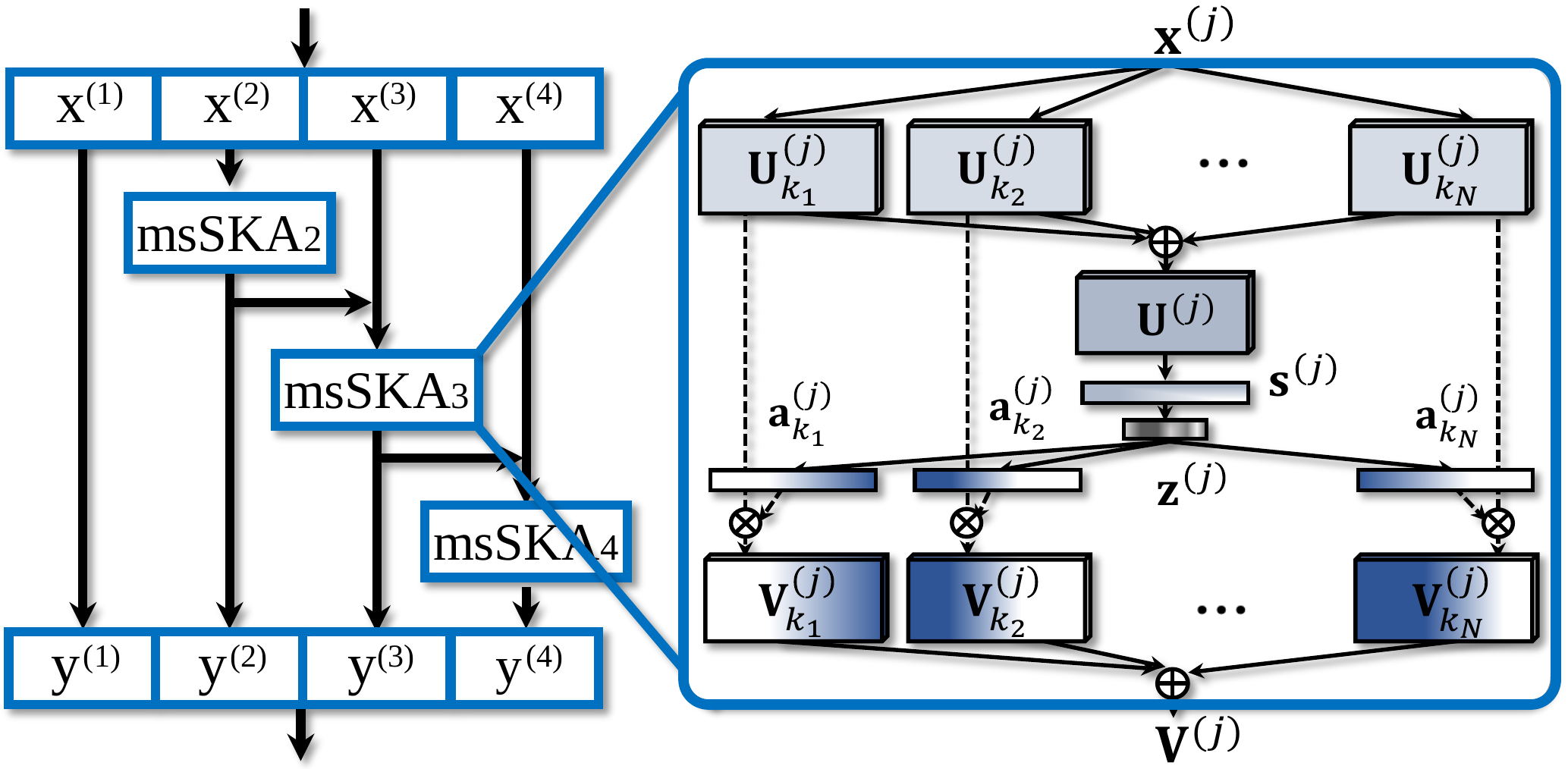}}
\vspace{0cm}
\end{minipage}
\centering
\caption{Multi-scale selective kernel attention (msSKA)}
\label{fig:msSKA}
\end{figure}

Compact feature $\textbf{z}$, attention weights $\textbf{a}$, and the output feature map $\textbf{V}$ is derived in the same manner with cwSKA. 
However, note that their dimensionalities are $F$, not $C$ because they operate upon the frequency dimension.
% Compact feature $z$ is derived in the same manner using FC, BN, and ReLU layers, i.e., $\textbf{z} = \text{ReLU}(\text{BN}(\textbf{W}\textbf{s}))$, where $\textbf{W} \in \mathbb{R}^{d \times F}$.
% Then the soft attention vector for $\textbf{U}_{k}$ is calculated as follows:
% \begin{equation}
  % \textbf{b}_{k,f} = {\exp(\textbf{B}_{k,f} \textbf{z}_{f}) \over {\sum^{K}_{l=1} \exp(\textbf{B}_{l,f}\textbf{z}_{f})}},
% \end{equation}
% \begin{equation}
  % \textbf{V}_{f} = \sum_{k=1}^{K} b_{k,f} \textbf{U}_{k,f}, \>\>\>
  % \sum_{k=1}^{K} b_{k,f} = 1,
% \end{equation}
% where $\textbf{B}_{k} \in \mathbb{R}^{F\times d_{f}}$ is the FC weight and $\textbf{b}_{k} \in \mathbb{R}^{F \times 1}$ is the frequency-wise soft attention vector.
% $b_{k,f}$ is the $f$-th element of $\textbf{b}_{k}$ and $\textbf{B}_{k,f} \in \mathbb{R}^{1 \times d_{f}}$ denote the $f$-th row vector of $\textbf{B}_{k}$.
% $\textbf{U}_{k,f}$ and $\textbf{V}_f \in \mathbb{R}^{C \times 1 \times T}$ are the $f$-th component of $\textbf{U}_{k}$ and $\textbf{V}$, respectively.
% We refer to the frequency-wise SKA as fwSKA.

\begin{figure*}[t!]
\begin{minipage}[b]{1.0\linewidth}
  \centering
  \centerline{\includegraphics[width=0.95\linewidth]{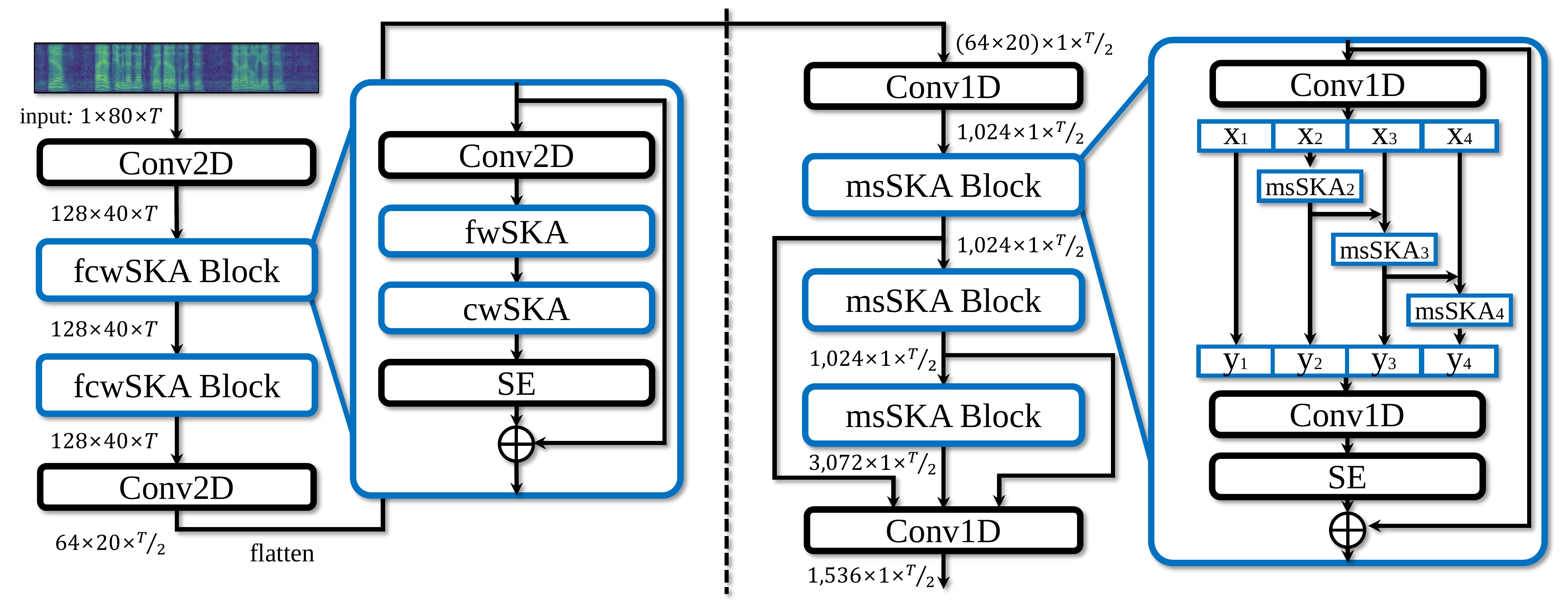}}
\vspace{0cm}
\end{minipage}
\centering\caption{The overall proposed architecture: The frequency-channel-wise SKA block-based front network (left) and the multi-scale SKA block-based TDNN network (right). This architecture is referred to SKA-TDNN.}
\label{fig:architecture}
\end{figure*}

\subsection{Multi-scale SKA (msSKA)}
\label{ssec:msSKA}
% The two variants of the SKA mechanism target 2D-CNNs, whereas we believe that the SKA mechanism can also be helpful for a module that composes residual backbone blocks, enhancing the module to process multi-scale data more efficiently. 
We apply both the conventional cwSKA and our proposed fwSKA to the 2D-CNN module that places in front of the ECAPA-TDNN. 
However, we believe that the SKA technique can be also complementary with the Res2net backbone block of the ECAPA-TDNN, which also focuses on processing data in a multi-scale fashion. 
Hence, the multi-scale SKA (msSKA) module is proposed.
%The msSKA module replaces the Res2net-based backbone module of the ECAPA-TDNN architecture.
The msSKA module replaces the single kernel 1D-CNN in Res2net block of the ECAPA-TDNN architecture. % 수정 2022.07.21.

In the msSKA module, a 2D input feature map $\textbf{X}\in \mathbb{R}^{C'\times T'}$ is first evenly divided into $s$ feature map subsets $\{\textbf{x}^{(j)}\}^{s}_{j=1} \in \mathbb{R}^{(C'/s) \times T'}$ where $s$ denotes the number of scales.
Let $\mathcal{G}_{k_{i}}:\textbf{x}^{(j)} \rightarrow \textbf{U}^{(j)}_{k_i} \in \mathbb{R}^{(C/s) \times T}$ be 1D convolution operator with kernel size $k_{i}$.
The 1D convolution $\mathcal{G}_{k_{i}}$ is applied to the each feature map subset $\{\textbf{x}^{(j)}\}^{s}_{j=1}$, and the output brunches are integrated by the an element-wise summation: %as follows:
\begin{equation}
  \textbf{s}^{(j)} = {1 \over {T}}\sum_{t=1}^{T}\textbf{U}^{(j)}(t), \>\>\>
  \textbf{U}^{(j)} = \sum_{i=1}^{N} \textbf{U}^{(j)}_{k_{i}}, 
\end{equation}
% \begin{equation}
%   \textbf{a}_{k}^{(j)} = {\exp(\textbf{A}_{k}^{i} \textbf{z}_{m}^{i}) \over {\sum^{K}_{l=1} \exp(\textbf{C}_{l,m}^{i}\textbf{z}_{m}^{i})}},
% \end{equation}
% \begin{equation}
%   \textbf{V}_{m}^{i} = \sum_{k=1}^{K} c_{k,m}^{i} \textbf{U}_{k,m}^{i}, \>\>\>
%   \sum_{k=1}^{K} c_{k,m}^{i} = 1,
% \end{equation}
where $\textbf{U}^{(j)}\in\mathbb{R}^{(C/s)\times T}$ is the $j$-th scale's fused feature map obtained via 1D-CNN with different kernel sizes.
%$\textbf{C}_{k,m}^{i}$, $c_{k,m}^{i}$, $\textbf{U}_{k,m}^{i}$, and $\textbf{V}_{m}^{i}$ are the $m$-th component of $\textbf{C}_{k}^{i} \in \mathbb{R}^{C/s\times d_{f}}$, $\textbf{c}_{k}^{i} \in \mathbb{R}^{C/s \times 1}$, $\textbf{U}_{k}^{i}$, and $\textbf{V}^{i}\in \mathbb{R}^{C/s\times F \times T}$, respectively.
Also, the $j$-th compact feature $\textbf{z}^{(j)}$, attention weights $\textbf{a}^{(j)}$, and output feature map $\textbf{V}^{(j)}$ are calculated as in the Section~\ref{ssec:cwSKA} and \ref{ssec:fwSKA}.
Finally, all $j$ feature maps are concatenated in the channel axis.

\section{Model architectures}
\label{ssec:model}
Figure~\ref{fig:architecture} illustrates the overall scheme of the proposed architecture.
%We propose three modules leveraging the SKA mechanism described in the previous Section, namely, fcwSKA, fwSKA, and msSKA block.
We propose three blocks leveraging the SKA mechanism described in the previous Section, namely, fcwSKA, fwSKA, and msSKA blocks. % 용어 통일 2022.07.19.
The fcwSKA stands for applying fwSKA and cwSKA in sequence.
%Except for the proposed msSKA module, all modules exist within the 2D-CNN module in front of the ECAPA-TDNN architecture.
%The msSKA module replaces the Res2net block within the ECAPA-TDNN block. 
Except for the proposed msSKA block, all blocks exist within the 2D-CNN block in front of the ECAPA-TDNN architecture. % 용어 통일 2022.07.19.
The msSKA block replaces the Res2net block within the ECAPA-TDNN. % 용어 통일 2022.07.19.

% \newpara{The fcwSKA block.} comprises two 2D-CNNs and two proposed fcwSKA blocks in front of the ECAPA-TDNN.
% Each fcwSKA block includes a 2D-CNN, fwSKA, cwSKA, SE layers sequentially with the residual connection.
\newpara{The fcwSKA block.} comprises a 2D-CNN, fwSKA, cwSKA, SE layers sequentially with the residual connection.
Each fcwSKA block is included in front of the ECAPA-TDNN. % 용어 통일 2022.07.19.

\newpara{The fwSKA block.} has the same architecture as the fcwSKA block, but only consists of the fwSKA layer between the 2D-CNN and SE layers. % 용어 통일 2022.07.19.

\newpara{The msSKA block.} resembles a typical Res2net backbone block within the ECAPA-TDNN. However, we adopt msSKA to each {\em scale} except one. % 용어 통일 2022.07.19.

\begin{table*}[t!]
\aboverulesep=0ex
\belowrulesep=0.75ex
\centering
\caption{The experimental results on the VoxCeleb1-O, VoxCeleb-E and VoxCeleb-H evaluation protocols. COS: Vanilla cosine similarity. TTA: Test time augmentation. SN: Adaptive score normalisation. $^\dagger$: Our implementation.}
\label{tab3:table}
\renewcommand{\tabcolsep}{1.20mm} % 0.75 vs 1.2
\scriptsize{
%\footnotesize{
%\small{
%\begin{tabular}{cc |ccc|ccc |ccc|ccc |ccc|ccc}
\begin{tabularx}{\linewidth}{ccc |ccc|ccc |ccc|ccc |ccc|ccc}
\specialrule{1.0pt}{0pt}{2.0pt}
    \multicolumn{1}{c}{\multirow{1}{*}{\textbf{}}}
    & \multicolumn{1}{c}{\multirow{1}{*}{\textbf{}}}
    & \multicolumn{1}{c}{\multirow{1}{*}{\textbf{}}}
    & \multicolumn{6}{c}{\multirow{1}{*}{\textbf{VoxCeleb1-O}}}
    & \multicolumn{6}{c}{\multirow{1}{*}{\textbf{VoxCeleb1-E}}}
    & \multicolumn{6}{c}{\multirow{1}{*}{\textbf{VoxCeleb1-H}}}   \\
        \arrayrulecolor{black!100}\cmidrule(lr){4-9} \cmidrule(lr){10-15} \cmidrule(lr){16-21}
        \multicolumn{1}{c}{\multirow{1}{*}{\textbf{Model}}}
        & \multicolumn{1}{c}{\textbf{Params}}
        &
        & \multicolumn{3}{c}{\multirow{1}{*}{\textbf{EER(\%)}}}
        & \multicolumn{3}{c}{\multirow{1}{*}{\textbf{MinDCF}}}
        & \multicolumn{3}{c}{\multirow{1}{*}{\textbf{EER(\%)}}}
        & \multicolumn{3}{c}{\multirow{1}{*}{\textbf{MinDCF}}}
        & \multicolumn{3}{c}{\multirow{1}{*}{\textbf{EER(\%)}}}
        & \multicolumn{3}{c}{\multirow{1}{*}{\textbf{MinDCF}}} \\
             
             \cmidrule(lr){4-6} \cmidrule(lr){7-9} \cmidrule(lr){10-12} \cmidrule(lr){13-15} \cmidrule(lr){16-18} \cmidrule(lr){19-21}
               &  \multicolumn{1}{c}{}     &        &  \textbf{Full}   & \textbf{3.0s}   & \multicolumn{1}{c}{\textbf{1.5s}}   &  \textbf{Full}   & \textbf{3.0s}   & \multicolumn{1}{c}{\textbf{1.5s}}  
                                           &  \textbf{Full}   & \textbf{3.0s}   & \multicolumn{1}{c}{\textbf{1.5s}}   &  \textbf{Full}   & \textbf{3.0s}   & \multicolumn{1}{c}{\textbf{1.5s}}  
                                           &  \textbf{Full}   & \textbf{3.0s}   & \multicolumn{1}{c}{\textbf{1.5s}}   &  \textbf{Full}   & \textbf{3.0s}   & \textbf{1.5s} \\        
\specialrule{0.8pt}{0pt}{2.0pt}
    \multirow{3}{*}{\makecell[c]{ResNet34 Q/SAP} \cite{heo2020clova}}   &       & \tiny{COS}     &  2.27   & 2.74   & 4.70   & 0.169   & 0.217   & 0.336   
                                                            &  2.33   & 2.90   & 4.60   & 0.169   & 0.216   & 0.334   
                                                            &  4.50   & 5.55   & 8.43   & 0.281   & 0.352   & 0.499 \\
    
                                                            &  \scriptsize{1.4M}   & \tiny{TTA}     &  2.23   & -      & -      & 0.167   & -       & -   
                                                            &  2.27   & -      & -      & 0.166   & -       & -   
                                                            &  4.37   & -      & -      & 0.283   & -       & -     \\
                                                            &       & \tiny{SN}     &  2.08   & 2.65   & 4.51   & 0.163   & 0.210   & 0.320   
                                                            &  2.18   & 2.79   & 4.49   & 0.151   & 0.201   & 0.313   
                                                            &  4.23   & 5.42   & 8.35   & 0.248   & 0.321   & 0.465 \\
    \arrayrulecolor{black!40}
    \specialrule{0.3pt}{0pt}{2.0pt}

    \multirow{3}{*}{\makecell[c]{ResNet34 H/ASP} \cite{heo2020clova}}  &        & \tiny{COS}     &  1.09   & 1.47   & 2.67   & 0.091   & 0.112   & 0.200   
                                                            &  1.28   & 1.59   & 2.64   & 0.094   & 0.114   & 0.184   
                                                            &  2.29   & 2.95   & 4.68   & 0.167   & 0.200   & 0.293 \\
    
                                                            &  \scriptsize{7.7M}     & \tiny{TTA}     &  1.06   & -      & -      & 0.083   & -       & -   
                                                            &  1.23   & -      & -      & 0.087   & -       & -   
                                                            &  2.23   & -      & -      & 0.155   & -       & -    \\
                                                            &       & \tiny{SN}   &  1.01   & 1.31   & 2.55   & 0.080   & 0.108   & 0.193   
                                                            &  1.14   & 1.46   & 2.51   & 0.080   & 0.103   & 0.171    
                                                            &  2.55   & 2.96   & 4.78   & 0.133   & 0.178   & 0.272  \\
    \arrayrulecolor{black!40}
    \specialrule{0.3pt}{0pt}{2.0pt}

    \multirow{3}{*}{\makecell[c]{ECAPA-TDNN$^\dagger$} \cite{desplanques2020ecapa}}     &         & \tiny{COS}     & 1.01    & 1.43   & 2.77   & 0.081   & 0.110   & 0.197  
                                                            & 1.21    & 1.52   & 2.75   & 0.088   & 0.105   & 0.183     
                                                            & 2.23    & 2.89   & 4.72   & 0.156   & 0.201   & 0.296   \\
    
                                                            & \scriptsize{14.7M}  & \tiny{TTA}    & 0.97    & -      & -      & 0.078   & -       & -   
                                                            & 1.19    & -      & -      & 0.086   & -       & -   
                                                            & 2.14    & -      & -      & 0.150   & -       & -    \\
                                                            &       & \tiny{SN}    & 0.96    & 1.35   & 2.59   & 0.077   & 0.105   & 0.188   
                                                            & 1.16    & 1.48   & 2.60   & 0.079   & 0.101   & 0.179      
                                                            & 2.10    & 2.79   & 4.59   & 0.135   & 0.181   & 0.274        \\
    \arrayrulecolor{black!40}
    \specialrule{0.3pt}{0pt}{2.0pt}

    \multirow{3}{*}{\makecell[c]{ECAPA-CNN-TDNN$^\dagger$} \cite{thienpondt2021ecapacnn}}   &       & \tiny{COS}     & 0.94    & 1.21   & 2.32   & 0.063   & 0.095   & 0.172
                                                            & 1.07    & 1.39   & 2.36   & 0.074   & 0.096   & 0.159     
                                                            & 2.03    & 2.64   & 4.34   & 0.129   & 0.169   & 0.255   \\
    %\multirow{2}{*}{\makecell[c]{TDNN$^\dagger$}\cite{thienpondt2021ecapacnn}}
                                                            & \scriptsize{27.6M} & \tiny{TTA}    & 0.91    & -      & -      & 0.063   & -       & -   
                                                            & 1.06    & -      & -      & 0.071   & -       & -   
                                                            & 2.04    & -      & -      & 0.123   & -       & -      \\
                                                            &       & \tiny{SN}    & 0.88    & 1.20   & 2.26   & 0.060   & 0.094   & 0.168    
                                                            & 1.01    & 1.35   & 2.31   & 0.069   & 0.092   & 0.154   
                                                            & 1.93    & 2.52   & 4.11   & 0.115   & 0.160   & 0.247  \\
    \arrayrulecolor{black!40}
    \specialrule{0.3pt}{0pt}{2.0pt}

    \multirow{3}{*}{\makecell[c]{MFA-TDNN$^\dagger$} \cite{liu2022mfa}}&       & \tiny{COS}       & 0.90   & 1.18   & 2.28    & 0.064   & 0.096   & 0.169
                                                            & 1.05    & 1.36   & 2.33   & 0.073   & 0.097   & 0.161
                                                            & 2.00    & 2.62   & 4.29   & 0.132   & 0.165   & 0.252       \\
    
                                                            & \scriptsize{24.9M} & \tiny{TTA}    & 0.86    & -      & -      & 0.068   & -       & -   
                                                            & 1.03    & -      & -      & 0.070   & -       & -
                                                            & 2.02    & -      & -      & 0.130   & -       & -      \\
                                                            &       & \tiny{SN}    & 0.84   & 1.16   & 2.20   & 0.059   & 0.092   & 0.161
                                                            & 0.98    & 1.30   & 2.27   & 0.066   & 0.093   & 0.152        
                                                            & 1.89    & 2.48   & 3.98   & 0.119   & 0.158   & 0.243   \\
    \arrayrulecolor{black!40}
    \specialrule{0.3pt}{0pt}{2.0pt}

    \multirow{3}{*}{\makecell[c]{ECAPA-CNN-TDNN\\ \scriptsize{with cwSKA}}}   &      & \tiny{COS}  & 0.91 & 1.19 & 2.26 & 0.067 & 0.094 & 0.162
                                                                            & 1.05 & 1.34 & 2.30 & 0.072 & 0.095 & 0.155 
                                                                            & 1.97 & 2.52 & 4.15 & 0.124 & 0.162 & 0.253  \\
    
                        & \scriptsize{28.3M}  & \tiny{TTA} & 0.83 & -      & -     & 0.060 & -      & -   
                                                            & 0.99 & -      & -     & 0.068 & -      & -   
                                                            & 1.95 & -      & -     & 0.120 & -      & -   \\
                                       &      & \tiny{SN}  & 0.83 & 1.15  & 2.20   & 0.061 & 0.089  & 0.155       
                                                            & 0.97 & 1.29  & 2.23   & 0.066 & 0.086  & 0.149
                                                            & 1.91 & 2.45  & 3.95   & 0.117 & 0.154  & 0.245 \\
    \arrayrulecolor{black!100}
    \specialrule{0.8pt}{0pt}{2.0pt}
    \specialrule{0.8pt}{0pt}{2.0pt}
%%%%%%%%%%%%%%%%%%%%%%%%%%%%%%%%%%%%%%%%%%%%%%%%%%%%%%%%%%%%%%%%%%%%%%%%%%%%%%%%%%%%%%%%%%%%%%%%%%%%%%%    
%%%%%%%%%%%%%%%%%%%%%%%%%%%%%%%%%%%%%%%%%%%%%%%%%%%%%%%%%%%%%%%%%%%%%%%%%%%%%%%%%%%%%%%%%%%%%%%%%%%%%%%

    \multirow{3}{*}{\makecell[c]{ECAPA-TDNN\\ \scriptsize{with msSKA}}} &    & \tiny{COS}       & 0.97    & 1.29   & 2.45   & 0.074   & 0.107   & 0.181
                                                             & 1.13    & 1.47   & 2.52   & 0.076   & 0.099   & 0.168
                                                             & 2.12    & 2.76   & 4.49   & 0.153   & 0.186   & 0.268    \\
                     & \scriptsize{16.7M}   & \tiny{TTA}    & 0.95    & -      & -      & 0.076   & -       & -   
                                                             & 1.12    & -      & -      & 0.078   & -       & -   
                                                             & 2.10    & -      & -      & 0.148   & -       & -     \\
                                     &      & \tiny{SN}     & 0.92    & 1.28   & 2.41   & 0.072   & 0.099   & 0.175     
                                                             & 1.09    & 1.44   & 2.48   & 0.074   & 0.096   & 0.164     
                                                             & 2.01    & 2.65   & 4.38   & 0.130   & 0.175   & 0.265 \\
    \arrayrulecolor{black!40}
    \specialrule{0.3pt}{0pt}{2.0pt}

    \multirow{3}{*}{\makecell[c]{ECAPA-CNN-TDNN\\ \scriptsize{with fwSKA}}} &      & \tiny{COS}
                                                              & 0.90 & 1.19  & 2.19  & 0.060 & 0.088 & 0.163
                                                              & 1.01 & 1.31  & 2.25  & 0.073 & 0.095 & 0.153 
                                                              & 1.93 & 2.49  & 4.05  & 0.122 & 0.161 & 0.248 \\
    
                       & \scriptsize{28.3M}  & \tiny{TTA}    & 0.82 & -     & -     & 0.060 & -     & -   
                                                              & 0.97 & -     & -     & 0.069 & -     & -   
                                                              & 1.90 & -     & -     & 0.115 & -     & - \\
                                       &      & \tiny{SN}    & 0.80 & 1.11  & 2.09  & 0.057 & 0.086 & 0.151
                                                              & 0.96 & 1.25  & 2.15  & 0.063 & 0.085 & 0.147      
                                                              & 1.86 & 2.38  & 3.87  & 0.111 & 0.148 & 0.239 \\
    \arrayrulecolor{black!40}
    \specialrule{0.3pt}{0pt}{2.0pt}

    \multirow{3}{*}{\makecell[c]{ECAPA-CNN-TDNN\\ \scriptsize{with fcwSKA}}} &       & \tiny{COS}     & 0.87    & 1.18   & 2.20    & 0.059   & 0.084   & 0.160     
                                                            & 0.99    & 1.29   & 2.19   & 0.069   & 0.091   & 0.148
                                                            & 1.90    & 2.44   & 4.00   & 0.118   & 0.154   & 0.246  \\
                      & \scriptsize{29.4M} & \tiny{TTA}    & 0.84    & -      & -      & 0.055   & -       & -   
                                                            & 0.98    & -      & -      & 0.067   & -       & -   
                                                            & 1.89    & -      & -      & 0.114   & -       & -    \\
                                    &       & \tiny{SN}    & 0.80    & 1.14   & 2.07   & 0.057   & 0.080   & 0.152     
                                                            & 0.93    & 1.20   & 2.06   & 0.061   & 0.084   & 0.137     
                                                            & 1.77    & 2.31   & 3.79   & 0.104   & 0.143   & 0.232 \\
    \arrayrulecolor{black!40}
    \specialrule{0.3pt}{0pt}{2.0pt}

    \multirow{3}{*}{\makecell[c]{\textbf{SKA-TDNN}}}  &       & \tiny{\textbf{COS}}     & \textbf{0.85}    & \textbf{1.14}   & \textbf{2.14}   & \textbf{0.054}   & \textbf{0.082}   & \textbf{0.154}     
                                                            & \textbf{0.97}    & \textbf{1.25}   & \textbf{2.12}   & \textbf{0.065}   & \textbf{0.087}   & \textbf{0.144}
                                                            & \textbf{1.87}    & \textbf{2.41}   & \textbf{3.95}   & \textbf{0.114}   & \textbf{0.150}   & \textbf{0.241}  \\

                                                            & \scriptsize{34.9M} & \tiny{\textbf{TTA}}    & \textbf{0.83}    & -      & -      & \textbf{0.053}   & -       & -   
                                                            & \textbf{0.94}    & -      & -      & \textbf{0.063}   & -       & -   
                                                            & \textbf{1.85}    & -      & -      & \textbf{0.111}   & -       & -    \\                                                           
                                                            &       & \tiny{\textbf{SN}}    & \textbf{0.78}    & \textbf{1.10}   & \textbf{2.05}   & \textbf{0.047}   & \textbf{0.078}   & \textbf{0.147}  
                                                            & \textbf{0.90}    & \textbf{1.18}   & \textbf{2.03}   & \textbf{0.059}   & \textbf{0.081}   & \textbf{0.134}     
                                                            & \textbf{1.74}    & \textbf{2.28}   & \textbf{3.77}   & \textbf{0.102}   & \textbf{0.138}   & \textbf{0.224}  \\
\arrayrulecolor{black!100}
\specialrule{1.0pt}{0pt}{2.0pt}
%\end{tabular}}
\end{tabularx}}
\end{table*}

\vspace{3pt}
% 용어 통일 2022.07.19. [networks / blocks / modules]
By applying above blocks to the ECAPA-TDNN or ECAPA-CNN-TDNN architecture, we propose four systems:
\begin{itemize}[nosep]
\item \newpara{ECAPA-TDNN with msSKA.} does not employ a 2D CNN-based block in front of the ECAPA-TDNN. It replaces the backbone blocks with the msSKA-based blocks (Figure~\ref{fig:architecture}, right).
In each msSKA block, we use $N=2$ for the 1D-CNNs with kernel sizes, where their sizes are 3 and 5.
Both the dilation and group size are set to 1, and the reduction ratio $C/d$ is 8.
We adopt a channel of 1,024 and a scale of 8.
\item \newpara{ECAPA-CNN-TDNN with fcwSKA.} places the proposed fcwSKA-based blocks instead of the front network of standard ECAPA-CNN-TDNN (Figure~\ref{fig:architecture} left).
In each fcw SKA block, the 2D-CNNs with the kernel sizes of 3$\times$3 and 5$\times$5 are exploited.
The dilation, the group size, and the reduction ratio are set to the same values as in the ECAPA-TDNN with msSKA.
For the multi-scale TDNN networks, a channel of 1,024 and a scale of 8 are used.
\item \newpara{ECAPA-CNN-TDNN with fwSKA.} has the same structure as the ECAPA-CNN-TDNN with fcwSKA, except for only containing the fwSKA layer in the SKA blocks.
\item \newpara{SKA-TDNN.} consists of both the fcwSKA block-based front and the msSKA block-based TDNN networks (Figure~\ref{fig:architecture}). 
We set the hyper-parameters to the same values used in the ECAPA-TDNN with msSKA and the ECAPA-CNN-TDNN with fcwSKA.
\end{itemize}

\vspace{3pt}
We adopt the channel and context-dependent statistic pooling \cite{desplanques2020ecapa} to aggregate the frame-level output features in all systems.
We adopt the equal-weighted summation of the additive angular margin (AAM) softmax \cite{deng2019aamsoft} and the angular prototypical (AP) \cite{chung2020in} objective functions to train all networks.

\section{Experiments}
\label{ssec:exp}
\subsection{Baseline model architectures}
We utilise the following six baselines: ResNet34 Q/SAP, H/ASP \cite{heo2020clova}, ECAPA-TDNN~\cite{desplanques2020ecapa}, ECAPA-CNN-TDNN~\cite{thienpondt2021ecapacnn}, MFA-TDNN \cite{liu2022mfa}, and the ECAPA-CNN-TDNN with cwSKA.
Among the baselines MFA-TDNN and ECAPA-CNN-TDNN with cwSKA would be the most competitive systems. 
MFA-TDNN is the most recently proposed system in this line of speaker verification research and we designed ECAPA-CNN-TDNN with cwSKA to validate how well the model performs when the conventional cwSKA is applied.
% Except for ResNet34 Q/SAP and H/ASP baselines where we used the pre-trained weight parameters, we re-implemented the remaining baselines.
Except for ResNet34 Q/SAP and H/ASP baselines where we used the pre-trained weight parameters, our own implementations were carried out for the other models. % 수정 2022.07.21.

\subsection{Dataset and evaluation protocol}
We use the development set of VoxCeleb2 dataset \cite{vox2} for training the models, which consists of 1,092,009 utterances from 5,994 speakers.
The evaluation is performed using VoxCeleb1 dataset~\cite{vox1} where we report the equal error rate (EER) and the minimum detection cost function (MinDCF) for three different evaluation protocols, namely, VoxCeleb1-O, VoxCeleb1-E and VoxCeleb1-H.
$P_{target}$=0.05 and $C_{miss}$=$C_{fa}$=1 are used to calculate the MinDCF metric.

% %%%%%%%%%%%%%%%%%%%%%%%%%%%%%%%%%%% FIGURE 2 %%%%%%%%%%%%%%%%%%%%%%%%%%%%%%%%%%%
\begin{figure*}[t!]
     \centering
     %\hfill
     \begin{subfigure}[b]{0.36\textwidth}
         \centering
        \flushleft
         \includegraphics[width=\textwidth]{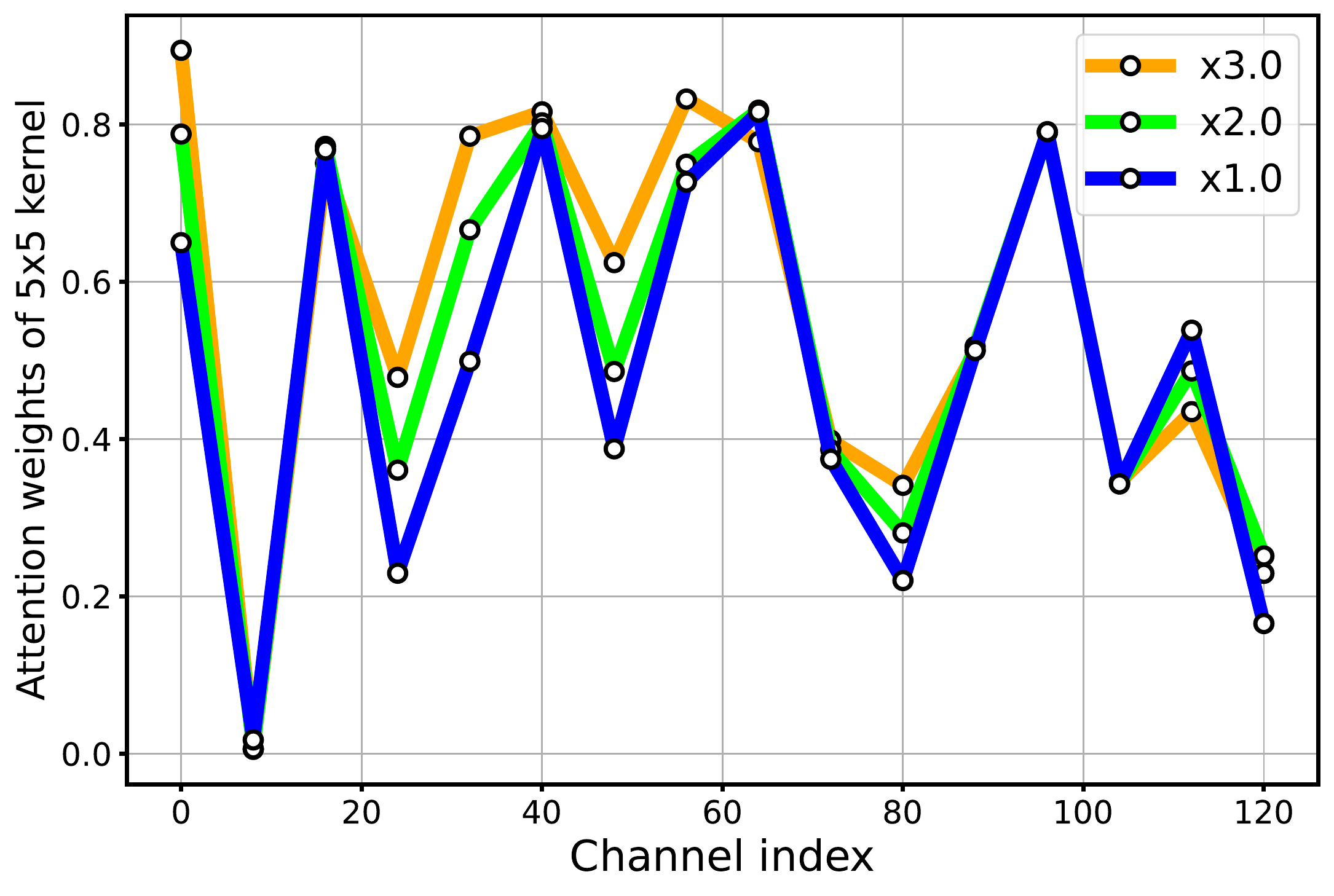}
         \vspace{-16pt}
     \end{subfigure}
     \hfill
     \begin{subfigure}[b]{0.36\textwidth}
         \centering
         \includegraphics[width=\textwidth]{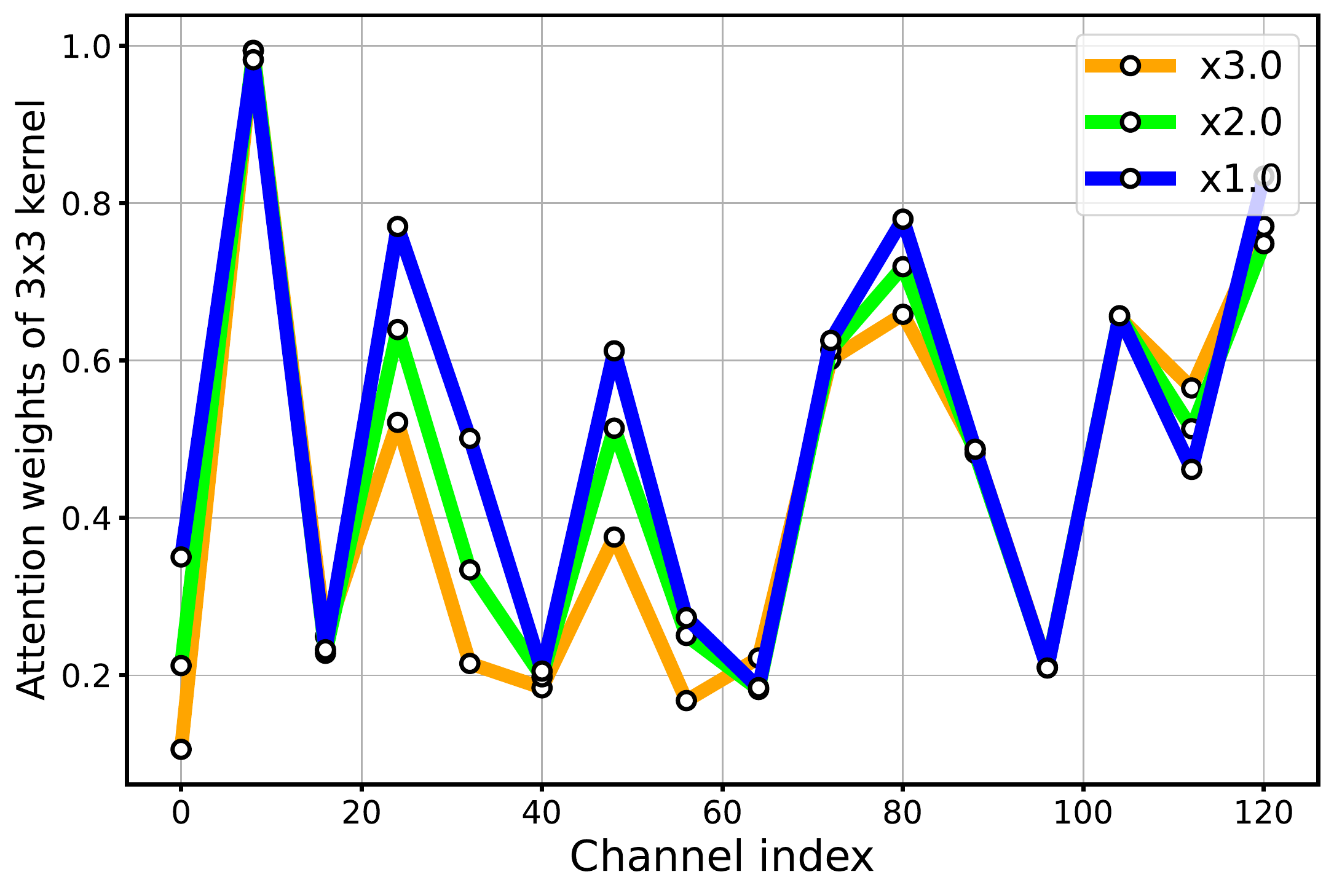}
         \vspace{-16pt}
     \end{subfigure}
     \hfill
     \begin{subfigure}[b]{0.27\textwidth}
         \centering
         \includegraphics[width=\textwidth]{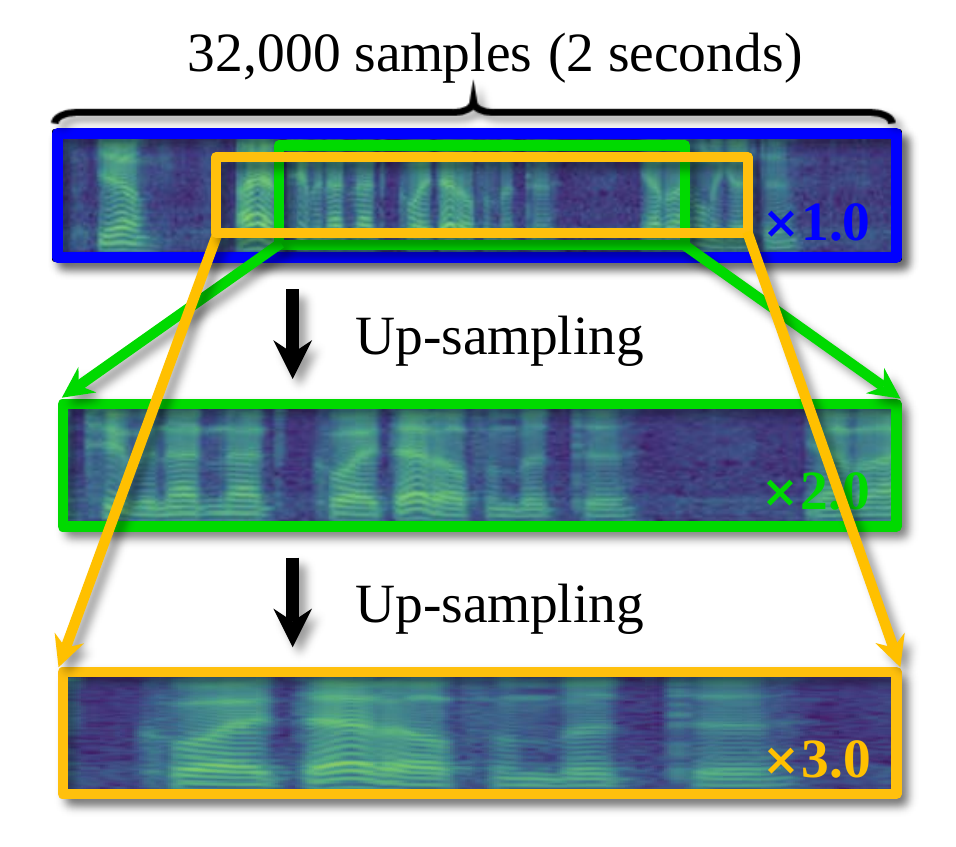}
         \vspace{-11pt}
     \end{subfigure}
\caption{The attention weights of 5$\times$5 (left) and 3$\times$3 (middle) kernels for each 128 channel index. The 3 input utterances with different resolutions (right). The input utterance is randomly sampled from VoxCeleb1 test set (\texttt{id10272/dkN2DIBrXqQ/00002.wav}).}
\label{fig:fig3}
\end{figure*}

\subsection{Back-end approaches for the scoring}
\label{ssec:scoring}
We report the performance of each model using three different back-end methods: (1) vanilla cosine similarity (COS), (2) test time augmentation (TTA), and (3) score normalisation (SN). % 수정 2022.07.21.
%We input the whole utterance at once for the vanilla cosine similarity and extract an embedding.
For the vanilla COS, the whole utterance is used as input to extract an embedding. % 수정 2022.07.21.
For the TTA~\cite{vox2}, we first segment each utterance into ten 4-second segments with overlaps.
The score for a given trial is derived by averaging cosine similarity values between all pairs of segments (i.e., 10$\times$10=100).
Finally, for the SN~\cite{matejka2017analysis}, we normalise the computed vanilla COS scores.
We adopt the VoxCeleb2 development set as the cohort set and then select the top 50,000 vanilla COS scores among cohort impostors to calculate the statistics for SN.

\subsection{Implementation details}
We implement models with the PyTorch library and conduct experiments using $4$ NVIDIA GeForce RTX 3090 GPUs in parallel\footnote{Implementation is available at \texttt {\url{https://github.com/msh9184/ska-tdnn.git}}.}
During training, we randomly crop an input utterance to a 2-second segment and then augment it with either MUSAN noises \cite{musan} or the simulated room impulse responses (RIRs) \cite{ko2017study}. 
Input features to the models are 80-dimensional log mel-filterbanks derived with a hamming window length of 25ms and hop-size of 10ms with 512-size FFT bins. 
We apply mean and variance normalisation to the log mel-filterbanks~\cite{ulyanov2016instance}.

The AAM-softmax objective function~\cite{deng2019aamsoft} adopts a margin of 0.2 and a scale of 30.
The AP objective function~\cite{chung2020in} uses one utterance for the prototype.
For training, all models are trained with a batch size of $200$ and optimised using an Adam optimiser \cite{kingma2014adam} with a weight decay of 2e-5.
The learning rate was scheduled via the cosine annealing with warm-up restart \cite{loshchilov2016sgdr} with a cycle size of 25 epochs, the maximum learning rate of 1e-3 and the decreasing rate of 0.8 for two cycles.

\section{Results}
\label{ssec:result}
\subsection{Main results}
Table \ref{tab3:table} describes the main experiments where we report the performances of several baselines and the proposed SKA-based models.
We additionally report the evaluation result on short duration scenario. 
Speaker embedding extracted from a full duration of an enrolment utterance is compared with either a test utterance with 3-second or 1.5-second duration.
We crop the middle part of an utterance to generate a short segment and if the utterance length is shorter than the target duration, we first duplicate and then perform cropping, following the protocol in \cite{jung2019RawNet, kim2021rawnext}.
We also report the results using the three scoring approaches, i.e., the vanilla COS, TTA, SN, described in Section \ref{ssec:scoring}. % changed 04.01.

By comparing rows 3 and 7 of Table \ref{tab3:table}, we observe marginal improvement by applying the proposed msSKA module in the backbone module (ECAPA-TDNN vs ECAPA-TDNN with msSKA).
Both models outperform the ResNet-based models (ResNet34 Q/SAP and H/ASP), which are commonly used in the SV field.
% 추가된 모델 성능 비교 부분 추가 2022.07.12.
%Next, we show that the use of the fcwSKA block-based front module (SKA-TDNN) boosts the performance compared to using the 2D-CNN or Res2Net-based front module (i.e., ECAPA-CNN-TDNN and MFA-TDNN).
Next, we show that the use of frequency-wise SKA in the front module (rows 8 and 9 of Table \ref{tab3:table}) helps improve the performance compared to the system using only conventional channel-wise SKA (rows 6 of Table \ref{tab3:table}).
%In addition, SKA-TDNN msSKA, including both fcwSKA block-based front module and msSKA block-based TDNN network, obtains further improved performance than the results of SKA-TDNN, achieving EER of 0.78\% and MinDCF of 0.047 on the VoxCeleb-O test set, respectively.
In addition, SKA-TDNN, including both fcwSKA block-based front module and msSKA block-based TDNN network, obtains the best performing result, achieving EER of 0.78\% and MinDCF of 0.047 on the VoxCeleb-O test set, respectively.
% 약간의 파라미터가 추가되긴 하지만, 성능을 높일 수 있는 부분 추가(?) 2022.07.12.
Although the SKA-based models introduce additional parameters, they effectively improve the performance without severely slowing down the inference process.

We also investigate the effect of SKA-based models on test utterances with different duration.
The proposed SKA-based models show consistent relative improvement under all test scenarios on different duration.
Compared to the SN results of best performing baseline, ECAPA-TDNN with cwSKA, on the VoxCeleb1-O test set, the SKA-TDNN obtains relative improvements of 4.35\% and 6.82\% in terms of EER on 3.0-second and 1.5-second test utterances, showing more improvement with shorter utterances.

Across all models including the proposed architectures, TTA and SN results show improved performance than those of typical cosine similarity (COS) where SN consistently showed the best performance on all evaluation sets.

\subsection{Analysis and interpretation}
We further design an additional experiment to gain insights on the SKA module's working mechanism. 
For this purpose, we observe the values of attention vectors ($\textbf{a}_{3\times3}, \textbf{a}_{5\times5}$) that decide on which kernel, either 3$\times$3 or 5$\times$5, is more utilised when utterances with different resolutions are input.
Utterances with different resolutions are generated using upsampling with interpolation as illustrated in the right side of Figure~\ref{fig:fig3}.

The left and the middle sides of Figure~\ref{fig:fig3} illustrate attention weights of 5$\times$5 kernel and 3$\times$3 kernel using an utterance randomly sampled from the VoxCeleb1 test set.
Through visualisation, we observe that for the 5$\times$5 kernel, attention values tend to increase as the input is upsampled more.
In contrast, for the 3$\times$3 kernel, attention values are the lowest when it is upsampled the most (yellow line).
% Comparing the original utterance (blue) with upsampled high resolution utterance (yellow), we observe that the attention weight for larger kernel becomes more activated.
We hence confirm that the SKA module adaptively selects the kernel size, thereby selecting the receptive field size, adjusted in a {\em data-driven} fashion.

\section{Conclusion}
\label{ssec:conclude}
This paper explored a selective kernel attention (SKA) module, allowing each convolutional layer to adaptively adjust kernel size based on an attention mechanism applied to both frequency and channel domain.
In addition, we proposed architectures by integrating the frequency-channel-wise SKA block-based front and the multi-scale SKA block-based TDNN networks.
Vast experiments conducted using three different evaluation protocols demonstrate that both proposed SKA-based modules boost the verification performance and applying both modules simultaneously performed the best. % changed 04.01
The SKA modules are relatively robust to short duration scenarios.

\section{Acknowledgements}
We thank Bong-Jin Lee, Hee-Soo Heo, Young-ki Kwon, and You Jin Kim at Naver Corporation for valuable discussions.

% \clearpage
\bibliographystyle{IEEEbib}
\bibliography{mybib}
\end{document}